\documentclass[12pt]{iopart}
\usepackage{textgreek}
 
\usepackage{braket}
\usepackage{xcolor}
\usepackage{setstack}
\usepackage{amstext}
\expandafter\let\csname equation*\endcsname\relax
\expandafter\let\csname endequation*\endcsname\relax
\usepackage{amssymb}
\usepackage{amsmath}
\usepackage{graphicx}

\begin{document}

\title{Polarization-gradient cooling of 1D and 2D ion Coulomb crystals}

\author {M. K. Joshi$^{1,2}$, A. Fabre$^{1}$\footnote{Present address: Laboratoire Kastler Brossel, Coll\`ege de France, CNRS, ENS-PSL University, Sorbonne Universit\'e, 11 Place Marcelin Berthelot, 75005 Paris, France}, C. Maier$^{1,2}$, T. Brydges$^{1,2}$, D. Kiesenhofer$^{1,2}$, H. Hainzer$^{1,2}$,   R. Blatt$^{1,2}$, C. F. Roos$^{1,2,*}$}

\address{$^1$Institute for Quantum Optics and Quantum Information, Technikerstra{\ss}e 21a, A-6020 Innsbruck, Austria}
\address{$^2$Institute for Experimental Physics, University of Innsbruck, Technikerstra{\ss}e 25, A-6020 Innsbruck, Austria}
\ead{$^{*}$christian.roos@uibk.ac.at}

\vspace{10pt}
\date{\today}

\begin{abstract}
We present experiments on polarization gradient cooling of Ca$^+$ multi-ion Coulomb crystals in a linear Paul trap. Polarization gradient cooling of the collective modes of motion whose eigenvectors have overlap with the symmetry axis of the trap is achieved by two counter-propagating laser beams with mutually orthogonal linear polarizations that are blue-detuned from the $S_{1/2}\leftrightarrow P_{1/2}$ transition. 
We demonstrate cooling of linear chains of up to 51 ions and 2D-crystals in zig-zag configuration with 22 ions. The cooling results are compared with numerical simulations and the predictions of a simple model of cooling in a moving polarization gradient. 
\end{abstract}


%
%
%
\section{Introduction}\label{sub:intro}
Laser cooling is an important technique to prepare neutral atoms and trapped ions at low temperature \cite{Wineland1979,Stenholm1986,Dalibard1989}. Doppler cooling \cite{Itano1982} enables the formation of ion Coulomb crystals \cite{Drewsen:2015,Thompson2015}, which are at the heart of many experiments on quantum computation, simulation, and precision measurements \cite{Bermudez:2017,Bruzewicz:2019,Blatt2012,Monroe:2019,Ludlow2015,Pruttivarasin:2015,Manovitz:2019}. In order to coherently manipulate ions with high fidelity, their motion in the trap should ideally be completely frozen out; hence, various laser cooling methods have been developed and demonstrated in the past three decades \cite{Eschner2003}. Resolved sideband cooling \cite{Diedrich1989,Roos1999}, Raman sideband cooling \cite{Monroe1995} and cooling by electromagnetically induced transparency (EIT) \cite{Morigi_2000,Roos2000a} are routinely used to prepare trapped ions in their motional ground state. 

While cooling to the ground state might be considered the ultimate goal of laser cooling, a cooling scheme should also be evaluated against other important metrics such as cooling speed, cooling range in frequency space, or the initial energy from which an ion can be cooled down to low temperatures. Conventional ground state cooling techniques, such as Raman and resolved-sideband cooling techniques are known to be well suited for achieving high ground state occupancy, but are in general slow in terms of cooling multiple modes \cite{Stutter2018}. On the other hand, the EIT cooling scheme, which is the fastest method of cooling to the ground state, maintains that cooling rate only within a limited range of motional frequencies and becomes slow when applied to low-frequency modes due to  the proximity of the bright state, which gets excited by sideband absorption, and the dark resonance, which suppresses carrier excitation.

The need for scaling up quantum information and quantum simulation experiments leads to new challenges in controlling both electronic and motional states of a large number of trapped ions. In trapped-ion clocks, the goal to increase the signal-to-noise ratio in a limited time interval leads to the development of multi-ion clocks, which require sub-Doppler cooling of ion strings to maintain their precision.

Large-size ion Coulomb crystals have been prepared by laser cooling \cite{Drewsen:2015,Thompson2015}, and sub-ensembles of vibrational modes even have been cooled below the Doppler limit in both Paul \cite{Lechner2016,Feng2020} and Penning traps \cite{Jordan2019}.
Yet, laser cooling schemes often under-perform when dealing with large crystals: a linear chain of ions can be created by operating a linear Paul trap at low axial confinement, which however gives rise to low-frequency collective motional modes with a strong occupation of high-lying phonon states. Planar crystals in Paul traps, which are of interest in the context of quantum simulation \cite{Porras2006,Richerme2016,Qiao:2020}, present a different kind of laser cooling challenge due to the limited number of laser beam directions for which the laser does not couple to the radio-frequency-driven ion motion. 
The quest for high-fidelity quantum operations in a trapped-ion-based quantum processor motivates us to explore novel approaches for cooling Coulomb crystals containing tens of ions.

Polarization gradient cooling (PGC) \cite{Dalibard1989} is a Sisyphus cooling technique, which has been widely used for cooling neutral atoms to very low temperatures \cite{Lett1988}. However, despite early theoretical works \cite{Wineland1992, Cirac1993a,Yoo:1993} investigating Sisyphus cooling of strongly bound atoms, i.e. atoms with wave functions localized to a length scale much smaller than the laser wavelength, the cooling scheme hasn't yet been extensively applied to cooling trapped ions. After an early demonstration \cite{Birkl1994}, polarization gradient cooling of trapped ions has been reported only recently in a regime where the ions were strongly bound to the trap \cite{Ejtemaee2017,Maslennikov:2019}. In experiments with up to four ions \cite{Ejtemaee2017}, the authors reported cooling ions to near the motional ground state from far outside the Lamb-Dicke regime (LDR) over a few hundred microseconds. In the current paper, we demonstrate polarization gradient cooling  of long ion strings and of planar ion Coulomb crystals to sub-Doppler temperatures.

Our results demonstrate that quantum computation, quantum simulation, and precision spectroscopy experiments can benefit from polarization gradient cooling. Specifically, for initializing the state of a trapped-ion quantum processor, many collective vibrational modes need to be cooled to low vibrational quantum numbers in order to avoid coupling strength fluctuations in quantum gates based on laser-ion interactions. Here, polarization gradient cooling can improve the gate fidelity without significantly increasing the length of the experimental duty cycle. Similarly, multi-ion clock experiments \cite{Keller:2019} require ion strings with as little secular energy as possible in order to reduce second-order Doppler shifts \cite{Chou2010,Brewer:2019}. In this context, polarization gradient cooling can be seen as a technique bridging the gap between Doppler cooling and ground-state cooling techniques targeting a small number of vibrational modes.

The current article is structured as follows. Section \ref{sub:theory} reviews and extends a simple model of polarization gradient cooling that predicts the achievable cooling limit and cooling rate. Our experimental apparatus and measurement techniques are described in section \ref{sub:apparatus}. In section \ref{sub:results}, we first present measurements carried out with a single trapped calcium ion before discussing results obtained with multiple ions in a 1D chain and a 2D crystal.

\section{ Theory of polarization gradient cooling}\label{sub:theory}
The theory of polarization gradient cooling of a bound atom has been developed previously \cite{Cirac1993a,Yoo:1993}, with a focus on  $j_g=1/2\leftrightarrow j_e=3/2$ transitions, with $j_g$ and $j_e$ being total angular quantum numbers for the ground and the excited electronic states. Here we will briefly present the main elements of the semiclassical theory developed in ref.~\cite{Cirac1993a} and derive the cooling rate and cooling limit for the case of a $j_g=1/2\leftrightarrow j_e=1/2$ transition.  

We consider an ion of mass $m$ confined in a harmonic potential with trap frequency $\omega_z$. The ion has a dipole-allowed transition connecting the $j_g=1/2$ ground state to an excited $j_e=1/2$ state. The transition is off-resonantly excited by two counter-propagating laser beams with linear-perpendicular-linear (lin-$\perp$-lin) polarizations. This arrangement of laser polarizations forms a periodically varying polarization gradient along the direction of the laser propagation \cite{Dalibard1989}. For simplicity, we assume that the cooling laser beams propagate in the direction of the quantization axis ($\hat{z}$) such that the ion interacts only with $\sigma$-polarized laser photons. 

For an ion moving along the $\hat{z}$ axis, its Zeeman ground states $|\pm\rangle=|j_g,m=\pm 1/2\rangle$ 
experience periodically varying ac-Stark shifts that are $\pi$ out of phase with respect to each other. In conjunction with the axial trapping potential $U_{trap}=\frac{1}{2} m\omega_z^2z^2$, these light shifts give rise to a state-dependent total potential energy given by 
\begin{equation}
    U_{\pm} = U_{trap}+\frac{1}{3}\Delta s \mp \frac{1}{3}\Delta s \sin(2kz+2\phi),\label{eq:lightshifts}
\end{equation}
where $k$ is the wave number, $\Delta$ the detuning from the dipole-allowed transition of the laser field and $\phi$ determines the position of the polarization gradient with respect to the trap center (for $\phi=0$, an ion at the trap center is driven by linearly polarized light). The intensity of the cooling beams is expressed in terms of the saturation parameter, $s=\frac{\Omega^2/2}{\Gamma^2/4 + \Delta^2}$, with $\Gamma$ the linewidth and $\Omega$ the Rabi frequency that the ion would experience on a $|+\rangle\leftrightarrow |j_e=3/2,m=3/2\rangle$ transition when interacting with only one of the two laser beams whose polarization was changed to being $\sigma_+$-polarized. In this case, the excited-state population would be given by $p=\frac{s/2}{1+s}$ for the definition of the Rabi frequency that we have chosen (i.~e., a $\pi$-pulse requiring a duration of $\tau=\pi/\Omega$).

\begin{figure}
    \centering
  \includegraphics[scale=1]{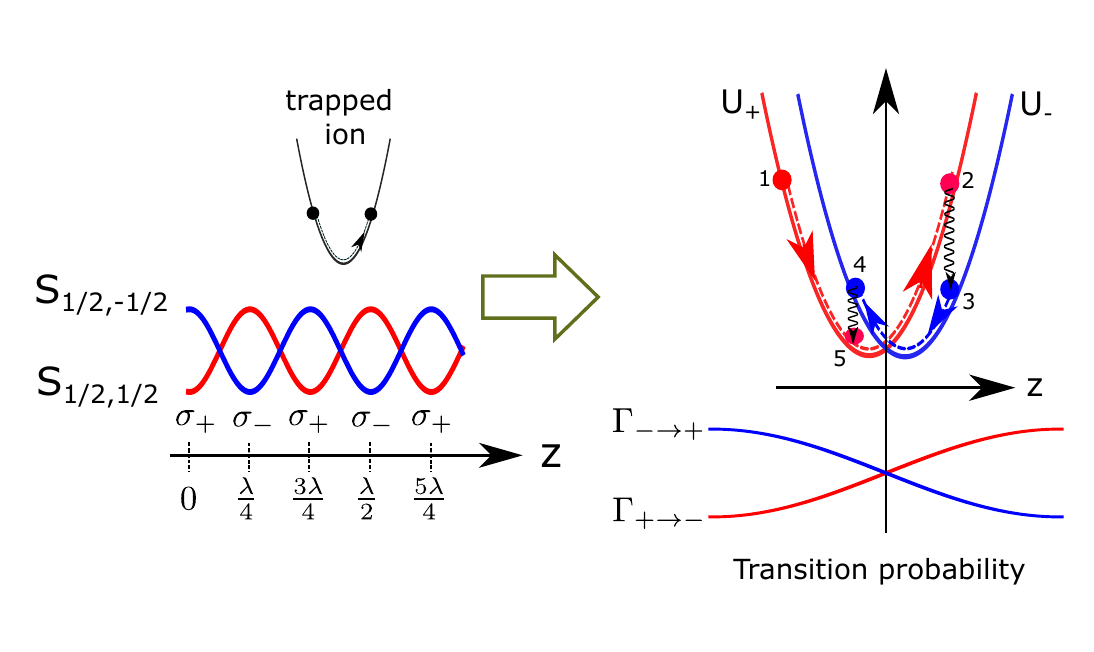}
 \caption{Principle of polarization gradient cooling. A trapped ion with an $S_{1/2}$ ground state in a polarization gradient coupling the two Zeeman ground states to an excited state. The polarization gradient makes the trapping potential state-dependent, by displacing it into opposite directions for the two Zeeman ground states. It also induces spatially varying pumping rates between the states. These two effects enable cooling a hot ion by favouring energy-reducing transitions (black squiggly arrows) over the reverse processes. An ion residing in one of these potential wells, for instance at step 1 or step 3, can experience energy-reducing transitions, i.e. at step 2 or step 4, while making harmonic oscillations about the potential well. These transitions between the two wells are favoured over the reverse transitions because of the spatially varying pumping rates between the states. The process can continue until the ion motion equilibrates to a state with minimum energy.}
    \label{fig:PGCIllustration}
\end{figure}

In addition to creating state-dependent potentials, the polarization gradient also gives rise to spatially dependent optical pumping rates between the ground states given by
\begin{equation}
     \Gamma_{\pm\rightarrow \mp}=\frac{1}{9}\Gamma s(1\mp\sin(2\phi))
     \label{eq:pumpingrates}.
\end{equation}
It is the interplay of spatially dependent light shifts and pumping rates that gives rise to an efficient cooling mechanism, sketched in Fig.~\ref{fig:PGCIllustration}. We assume that the potential of the ion trap is much stiffer than the potential of the polarization gradient so that the latter creates state-dependent potentials $U_\pm(z)$ for the two Zeeman ground states $|\pm\rangle$ that can be treated as being harmonic. If the potential minima of $U_\pm$ are spatially displaced, a hot ion can be cooled because the spatially varying pumping rate will favour state changes giving rise to an energy loss (as indicated in the figure) over the reverse processes. Note that for this picture to hold, the pumping rates $\Gamma_\pm$ have to be smaller than the ion's oscillation frequency. The ion motion will come to a steady state when the cooling is balanced by heating by spontaneously scattered photons and dipole force fluctuations, and additionally by scattering processes at random times in which the ion is pumped to the other Zeeman ground state. 

Following ref.~\cite{Cirac1993a}, we calculate the cooling limit and cooling rate by a simple model, in which an ion having a motional energy $E$ stochastically switches between its electronic ground states with rates $\Gamma_\pm$; the model is valid in the low saturation limit, where very little population resides in excited electronic states. By calculating the average motional energy change induced by transitions to the other Zeeman state and accounting for motional heating by absorption and emission processes, one obtains the following differential equation describing the cooling dynamics:
\begin{align} \label{eq:rate}
    \dot{E} & = p_+ \braket{\Gamma_{+ \rightarrow -}(z)(U_-(z)-U_+(z))}_+\\ \nonumber
    & + p_- \braket{\Gamma_{- \rightarrow +}(z)(U_+(z)-U_-(z))}_-  + H_{sc}\hbar\omega_z, 
\end{align}
where $p_\pm=\frac{1}{2}(1\pm\sin(2kz+2\phi))$ are the ground state populations in steady state, and $\braket{...}_+$ and $\braket{...}_-$ indicate the averages over the ion's spatial probability distribution  in the potential $U_+$ and $U_-$, respectively. The term $H_{sc}\hbar\omega_z$ accounts for the momentum diffusion due to fluctuations of the radiation pressure and optical dipole forces arising from the laser-ion interaction \cite{Ashkin1980}. In the LDR, the resulting heating is conveniently described in terms of carrier and sideband transitions. In this picture, two heating processes need to be considered: (1) absorption on the carrier transition followed by spontaneous emission on the upper or lower vibrational sideband and (2) absorption on the upper or lower sideband followed by spontaneous emission on the carrier transition. At $z=0$, these two processes lead to a heating rate $H_{sc}=H_{carr}+H_{sb}$ given by
\begin{align}
H_{carr}&=\frac{\alpha}{3}\eta^2\Gamma s(1-\sin^22\phi)\\
H_{sb}&=\frac{1}{3}\eta^2\Gamma s(1+\sin^22\phi),
\end{align}
where $\eta = \sqrt{\hbar k^2/(2 m \omega_z)}$  is the Lamb-Dicke parameter and $\alpha=1/3$ results from the spatially isotropic spontaneous emission on the $S_{1/2}\leftrightarrow P_{1/2}$ transition that we are considering. Note that the dependence on the phase $\phi$ of the polarization gradient differs for these processes as the carrier coupling is maximum in the anti-nodes of the standing waves whereas the sideband coupling is maximum in the nodes \cite{Cirac1992}. After expanding the pumping rates $\Gamma_\pm$ and the potentials $U_\pm$ to first order in $kz$, an evaluation of the first two terms of equation (\ref{eq:rate}) gives rise to a term $\sim -E$ cooling the ion if the polarization gradient is blue-detuned with respect to the atomic transition. Additionally, there is a heating term proportional to the energy that an ion at the minimum of $U_\pm$ gains when being excited to the other potential $U_\mp$ (and vice versa). This leads to 
\begin{equation}
    \dot {E} = -W(\phi)E + H(\phi)\hbar\omega_z,
\end{equation}
with cooling and heating rates 
\begin{align}
    W(\phi) &= \frac{16}{9}\eta^2\Gamma s\xi \cos^22\phi, \label{eq:coolingrate}\\
    H(\phi) &= \phantom{,}\frac{2}{9}\eta^2\Gamma s(8\xi^2\cos^4 2\phi+2+\sin^22\phi), \label{eq:heatingrate}
\end{align}
and $\xi=\frac{\Delta s}{3\omega_z}$. If the ion experiences perfectly circularly polarized light ($\phi=\pm\pi/4$), the cooling rate vanishes. The most efficient cooling occurs at $\phi=0$ when the ion is placed at the steepest slope of the optical potential. Then, the mean phonon number in steady state is given by
\begin{equation}
    \braket{n_0(\xi)} = \frac{H(0)}{W(0)}-\frac{1}{2} = \xi +\frac{1}{4\xi} -\frac{1}{2}.
    \label{eq:nbarVsXi}
\end{equation}
A minimum phonon number of $\min(\braket{n_0})=1/2$ is obtained for $\xi=\frac{1}{2}$, when the optical well depth ($2\Delta s/3$) becomes equal to the trap frequency.

Experimentally, it is very challenging to achieve the sub-wavelength stability required to position the ion at a precise phase of the polarization gradient. However, there are two ways of overcoming this problem:
\paragraph{Cooling of many-ion crystals.} When cooling motional modes of ion crystals containing many ions, the ions will roughly uniformly sample the phases of the polarization gradient so that the exact positioning of the polarization gradient no longer matters. One can then average the heating and cooling rates, eqs.~(\ref{eq:coolingrate}), (\ref{eq:heatingrate}), over $\phi$ in order to calculate the cooling limit for a collective vibrational mode with frequency $\omega_i$
\begin{equation}
    \langle n \rangle = \frac{\int H(\phi)d\phi }{\int W(\phi) d\phi}-\frac{1}{2} = \frac{3}{4}\xi+\frac{5}{8\xi}-\frac{1}{2},\label{eq:nbarphaseaveraged}
\end{equation}
with $\xi=\frac{\Delta s}{3\omega_i}$. The energy which is minimized for $\xi_{min}=\sqrt{5/6}\approx 0.91$, resulting in 
$\min{\langle n\rangle}=\sqrt{\frac{15}{8}}-\frac{1}{2}\approx 0.87$.
\paragraph{Cooling in a moving polarization gradient.}
For a single ion or a few-ion crystal, the cooling scheme can be made robust by means of a moving polarization gradient \cite{Wineland:1992} that can be created by introducing a small frequency detuning $\delta$ between the two counter-propagating beams. This detuning should be higher than the cooling rate $W$; otherwise the mean phonon number would adiabatically follow the steady-state value $\braket{n}_\phi$ which, on average, would lead to an increased motional energy and large shot-to-shot fluctuations caused by random values of $\phi$ at the end of the cooling pulse. On the other hand, $\delta<\omega_z$  is required to make the ion sample the potential well before the polarization gradient changes considerably. If $W<\delta<\omega_z$, eq.~(\ref{eq:nbarphaseaveraged}) predicts the steady-state mean phonon number.

As pointed out in ref.~\cite{Wineland:1992}, the semiclassical description of laser cooling of a harmonically trapped particle in the Lamb-Dicke regime is expected to give accurate results even for particles cooled close to the zero-point energy of the oscillator. We have validated the predictions of the simple cooling model described above by numerically solving the master equation describing polarization gradient cooling. The simulations, presented in the appendix of this paper, demonstrate that eqs.~(\ref{eq:coolingrate})-(\ref{eq:nbarphaseaveraged}) accurately describe the cooling of an ion deep in the LDR, i.e. where $\eta\sqrt{\langle n\rangle + 1/2}\ll 1$ holds.

\section{Experimental apparatus and measurement techniques}\label{sub:apparatus}
A linear Paul trap with blade-shaped RF electrodes \cite{Schmidt-Kaler2003} is used for confining both linear strings and planar crystals of $^{40}$Ca$^+$ ions. A detailed description of the experimental apparatus can be found in 
Ref.~\cite{Hempel2014}. 
A magnetic field of $B\approx 4.18$ G pointing along the trap's rf-zero line defines the ions' quantization axis and leads to a splitting of 11.714 MHz between the two Zeeman sublevels of the $S_{1/2}$ manifold. Ions are precooled using a Doppler cooling laser beam with an elliptical beam profile that illuminates the ion chain under an angle of $45$ degrees, and has an overlap with all collective modes of motion.

In the current article, we present experimental results of polarization gradient cooling along the axis of the ion chain ($\hat{z}$ axis). Holes through the trap's tip electrodes, with a diameter of 0.5 mm, allow two counter-propagating laser beams to pass through along $\hat{z}$, thus coupling to the axial motion only. 
Polarization gradient cooling is performed on the $S_{1/2} \leftrightarrow P_{1/2}$ transition at 397~nm. The laser used for Doppler cooling also provides the two counter-propagating beams that are blue-detuned by $\Delta=2\pi\times 210$~MHz from the cooling transition. The beams are linearly polarized with orthogonal polarizations, creating a polarization gradient that alternates between right circular and left circular polarizations along the direction of light propagation. The detuning of the beams can be controlled independently by means of two acousto-optical modulators. In most experiments, their frequency difference was set to $\delta=2 \pi \times 60$~kHz to create a polarization gradient moving at a rate greater than the cooling rate.
 
 We calibrate the intensity of the two cooling beams in the trap center separately with errors below 10\% by optically pumping the ions to one of the Zeeman ground states and subsequently measuring the equilibration time constant of the two ground state populations after one of the cooling beams is switched on. For this, Zeeman ground state populations are measured by transferring the respective population to the $D_{5/2}$ state with a narrow linewidth laser beam at 729~nm, which is also used for motional state analysis.
 
\section{Results and discussion}\label{sub:results}
We investigate polarization gradient cooling using the following experimental sequence: In the first step, calcium ions are prepared in the $S_{1/2}$ state and Doppler cooling is performed for 3 ms. Next, a polarization gradient cooling pulse is employed for up to 1 ms. To prevent optical pumping to the $D_{3/2}$ state by spontaneous decay of the $P_{1/2}$ state, a repumper at 866~nm is switched on during the Doppler and polarization gradient cooling steps. In the third step, ions are prepared in the $|\mbox{S}_{1/2}, m=1/2\rangle$ state by frequency-resolved optical pumping on the $|\mbox{S}_{1/2}, m=1/2\rangle \leftrightarrow |\mbox{D}_{5/2}, m=3/2\rangle$ transition, in conjunction with the 854 nm laser beam, preparing the ions in the $|\mbox{S}_{1/2}, m=1/2\rangle$ sublevel with a probability greater than 99.9\%. The final step involves mapping the information about the motional state onto the electronic states, via a 729 nm laser pulse on either a carrier or a first-order sideband of the $\mbox{S}_{1/2}\leftrightarrow\mbox{D}_{5/2}$ transition followed by quantum state detection via fluorescence detection at 397~nm. Spatially resolved measurements of the fluorescence of an ion crystal are recorded with a CCD camera in order to assign the quantum state of individual ions.  

\subsection{Cooling of a single ion}\label{sub:Single}
\begin{figure}
    \centering
     \includegraphics[scale=1.]{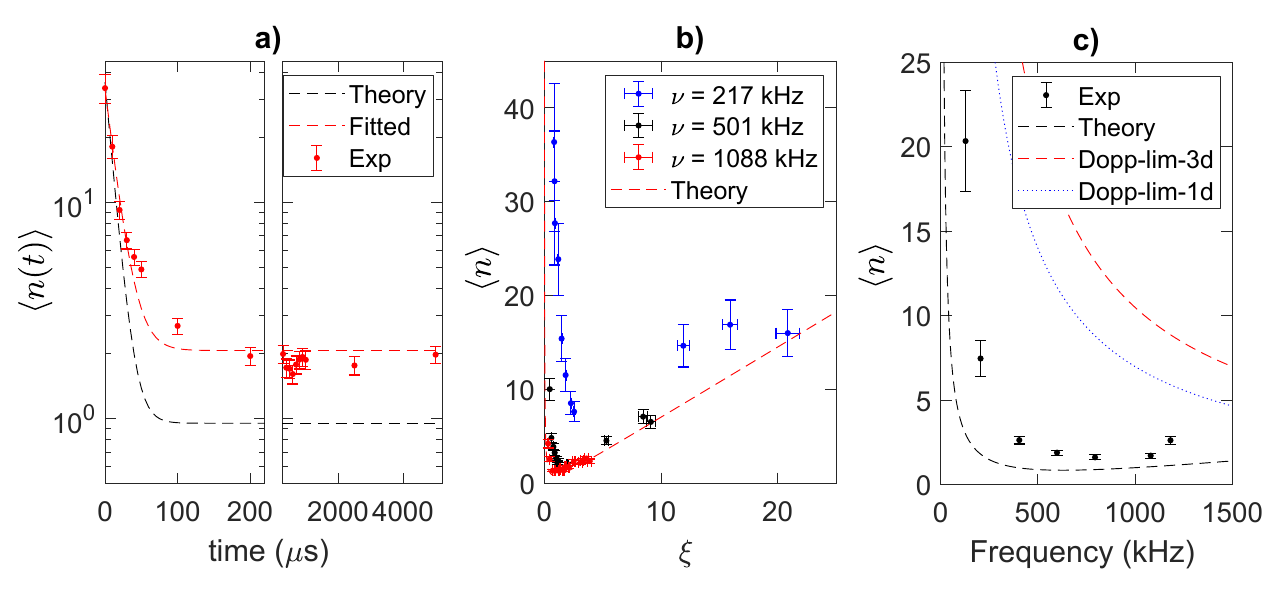}
      \caption{Cooling of a single ion. (a) Mean phonon number $\braket{n}$ as a function of cooling time at $\omega_z =2\pi\times 1.088$ MHz and 
      $\xi = 1.35(14)$. The simulated cooling dynamics (dashed black curve) for the experimental parameters is based on averaging eqs.~(\ref{eq:coolingrate}) and (\ref{eq:heatingrate}) over all phases $\phi$. The dashed red curve is a fit to the experimental data. (b) Mean phonon number $\braket{n}$ in steady state as a function of the normalized saturation parameter $\xi= \Delta s/3\omega_z$ at 1088~kHz (in red), 501~kHz (in black) and 217~kHz (in blue) axial trap frequencies, together with the cooling limit predicted by eq.~(\ref{eq:nbarVsXi}) (dashed line). (c) $\braket{n}$ as function of trap frequency $\omega_z$ at $s \sim 0.0085$ (15) , which is predicted to yield optimum cooling for $\omega_z = 2\pi\times 650$~kHz. The dashed black line is predicted by eq.~(\ref{eq:nbarVsXi}). The blue and red dashed lines represent the cooling limits of one- and three-dimensional Doppler cooling, respectively.}
    \label{fig:SingleIonMeasurement}
    
    
\end{figure}

We start by investigating polarization gradient cooling of a single ion. 
The cooling time constant is measured by applying cooling pulses of variable duration $\tau$ to a Doppler-cooled ion, followed by motional state analysis. For the latter, we drive carrier or first-order sideband Rabi oscillations, which we fit with a thermal phonon distribution. An example of cooling an ion is displayed in Fig.~\ref{fig:SingleIonMeasurement} (a), which shows that the ion motion equilibrates to a steady-state value that is a bit above the cooling limit predicted by eq.~(\ref{eq:nbarphaseaveraged}), and with a slightly lower rate. The resulting data is fitted with a function $f(\tau)=\langle n(\infty)\rangle + (\langle n(0)\rangle-\langle n(\infty)\rangle)\exp(-W\tau)$ in order to extract the cooling rate, where parameters $\langle n(0)\rangle$ and $\langle n(\infty)\rangle$ correspond to mean phonon numbers at $\tau=0$ and $\tau=\infty$, respectively. Here, for $\omega_z=2 \pi \times 1088$~kHz and $\xi=1.35(14)$, we find a cooling rate $W\approx66000$~s$^{-1}$. Taking into account the slightly non-exponential decay that we observe, the data shows that equilibrium is achieved after less than 200~$\mu$s.

For a cooling pulse of a duration $\tau$ much longer than the cooling time constant $1/W$, we measure the mean phonon number $\braket{n}$ in steady state  as a function of the trap frequency and the intensity of the polarization gradient.
Figure \ref{fig:SingleIonMeasurement} (b) shows the measured $\braket{n}$ as a function of the dimensionless parameter $\xi=\Delta s/3\omega_z$ for trap frequencies of 1088~kHz, 501~kHz, and 217~kHz, while varying the intensities of both laser beams to achieve the desired values of the saturation parameter $s$; please note that the assumption of low saturation holds in all experiments we carried out ($s<0.07$). For the highest trap frequency, optimum cooling is achieved for $\xi \sim 1$ in agreement with the prediction of eq.~(\ref{eq:nbarphaseaveraged}), which is indicated by a dashed red line. The measured mean phonon number is slightly above this limit (see also Fig.~\ref{fig:nbarvisXimovgrad} in the appendix showing the same data on an expanded scale). Master equation simulations confirm that this discrepancy can be explained by the non-zero Lamb-Dicke parameter of $\eta=0.17$. 

For the data taken at low trap frequency, the differences between measured and predicted mean phonon numbers become very pronounced, in particular at low values of $\xi$. For $\omega_z=2 \pi \times 217$~kHz, the minimum motional energy ($\langle n\rangle=7.7(1.1)$~phonons) is obtained for an optical potential with a depth that is significantly higher than the optimum value $\xi_{min}=\sqrt{5/6}$ predicted by eq.~(\ref{eq:nbarphaseaveraged}). This observation could be explained by electric field noise giving rise to motional heating at a rate that is much higher than the heating rate $H$ intrinsic to polarization gradient cooling. Alternatively it could be due to a reduction of the cooling rate outside the LDR. In our experiment, we observed a heating rate of 1350(85)~quanta/s at 217~kHz trap frequency. However, we infer from master equation simulations a cooling limit $\langle n\rangle>5$ phonons and estimate the cooling rate to be higher than 30000~quanta/s. Consequently, we conclude that electric field noise is not to blame and that the raised cooling limit is due to leaving the LDR.

Despite the increased cooling limit at very low trap frequencies, polarization gradient cooling is capable of cooling an ion below the Doppler limit (i.e. $\braket{n} = \frac{\Gamma}{4\omega_z}(1+\alpha)$ for optimum cooling in one dimension) over a wide range of trap frequencies. Figure~\ref{fig:SingleIonMeasurement} (c) shows a measurement of $\langle n\rangle$ as a function of the trap frequency, which ranged from 127~kHz up to 1200~kHz, with optimum cooling $\xi_{min}$ achieved for $\omega_z=2 \pi \times 650$~kHz. 
Here, sub-Doppler cooling is achieved over the entire frequency range.

\subsection{Cooling of 1D ion chains}\label{sub:1D}
In this section, we present experimental results of PGC of linear ion strings trapped in a potential with transverse trapping frequencies of $\omega_x = 2 \pi \times 2.67$ MHz and $\omega_y = 2 \pi \times 2.64$ MHz. The axial trap frequency was set to $\omega_z = 2 \pi \times 217$ kHz except for experiments with 51 ions, where it was lowered to $\omega_z = 2 \pi \times 127$ kHz.

\begin{figure}
    \centering
\includegraphics[scale=0.85]{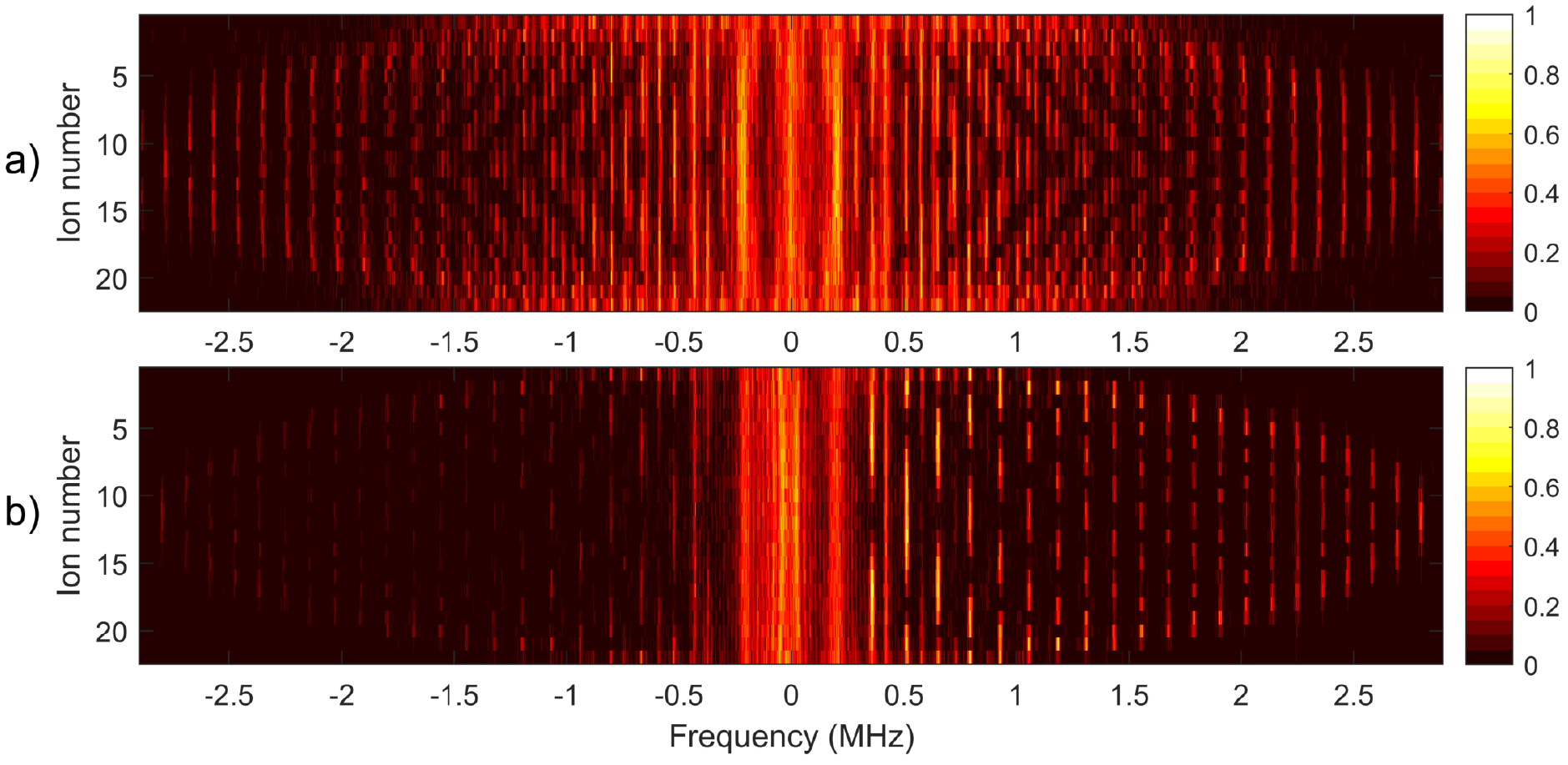}
    \caption{Spectra of a 22-ion chain on the $\mbox{S}_{1/2},m=1/2\leftrightarrow\mbox{D}_{5/2},m=5/2$ transition including its first-order axial sidebands after Doppler cooling (a) and polarization gradient cooling (b). The quantum state of individual ions is measured by recording their fluorescence with a CCD camera. The reduced motional energy after polarization gradient cooling leads to a weakening of the sideband transitions. }
    \label{fig:22ionspectrum1d}
\end{figure}

Spatially resolved sideband spectroscopy of a 22-ion string demonstrates the advantage of polarization gradient cooling over Doppler cooling as shown in  Fig.~\ref{fig:22ionspectrum1d} in a visually compelling way: a laser beam propagating along the symmetry axis of the ion trap is used to excite the $|\mbox{S}_{1/2}, m=1/2\rangle\leftrightarrow |\mbox{D}_{5/2}, m=3/2\rangle$ transition including all first-order axial sidebands. The figure shows spectra obtained after Doppler cooling, (a), and with an additional polarization gradient cooling pulse of 1~ms duration, (b). Clearly, the excitation strength on the sidebands is strongly reduced by polarization gradient cooling. Moreover, an asymmetry in the excitation strength $p_{red}$ ($p_{blue}$) between red (blue) sidebands appears, which, however, does not indicate ground state cooling for {\sl collectively} probed ion crystals \cite{Lechner2016} (This is in contrast to the single-ion case where cooling close to the ground state can be deduced from the relation $\langle n\rangle = p_{red}/(p_{blue}-p_{red})$ \cite{Leibfried:2003}, which is based on matching pairs of levels on the red and blue sideband transitions having the same coupling strength; for collectively excited multi-ion sideband transitions, there are no matching ladders of coupled levels). The spatially heterogeneous excitation probability on the sidebands seen in Fig.~\ref{fig:22ionspectrum1d} reflects the different Lamb-Dicke parameters the ions have \cite{James1998}. 

\begin{figure}[ht]
\centering
\includegraphics[scale=1]{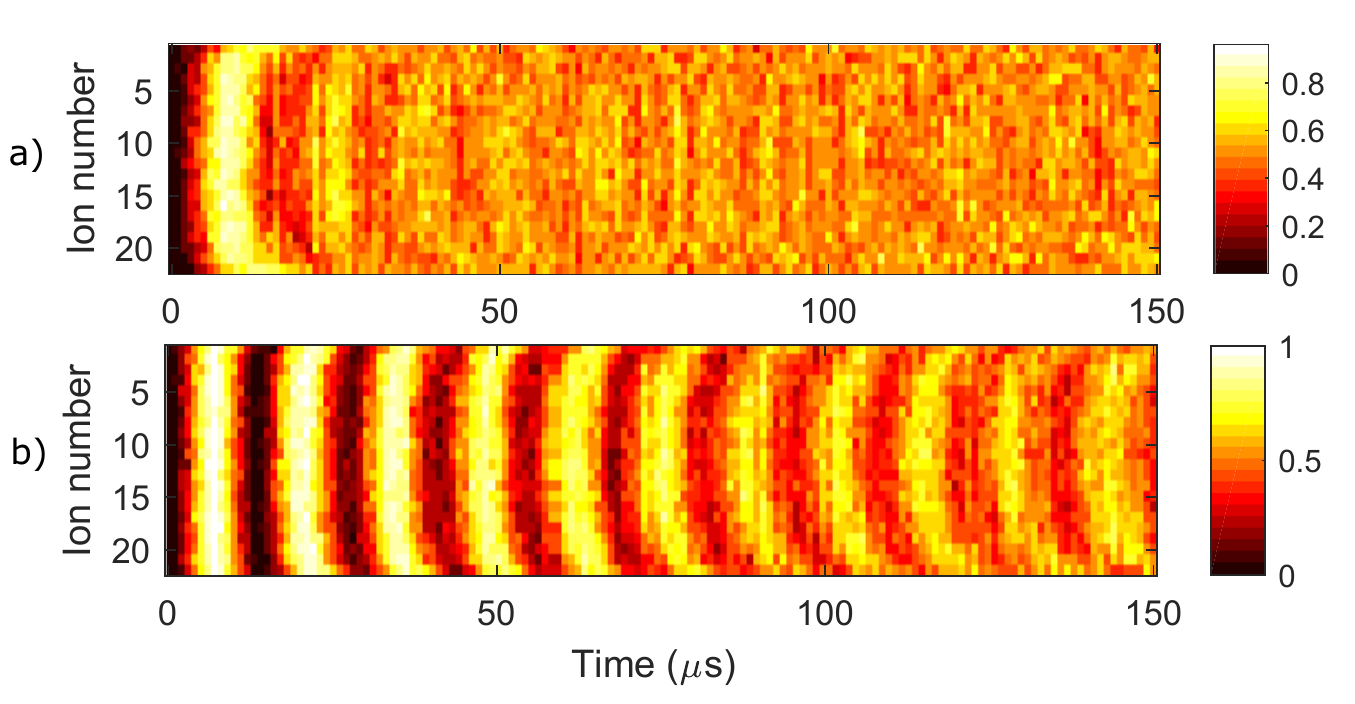}
\caption{Rabi oscillations of a 22-ion chain driven by a 729~nm laser beam propagating along the symmetry axis of the linear trap. Rabi oscillations of Doppler-cooled ions (a) damp out rapidly whereas the oscillations of polarization-gradient cooled ions are faster and damp out more slowly due to the stronger coupling and reduced shot-to-shot coupling strength fluctuations on the carrier transition. For a discussion of the spatial dependence of the average coupling strength, see main text.}
\label{fig:Rabiflops1d}
\end{figure}

Another way of globally assessing the influence of polarization gradient cooling on the motional state of the ion crystal is to drive carrier oscillations. The carrier coupling strength $\Omega_j$ of ion $j$ depends on the phonon number $n_k$ of mode $k$ as \cite{Leibfried:2003}
\begin{equation}
\Omega_j^{\mathbf{n}}=\Omega_{0,j}\prod_{k=1}^NL_{n_k}(\eta_{jk}^2),
\label{eq:RabiFrequency}
\end{equation} where $\mathbf{n}=(n_1,\ldots,n_N)$ is the vector of phonon numbers,  $L_n(\eta^2)$ a Laguerre polynomial, $\eta_{jk}$ the Lamb-Dicke parameters, and $\Omega_{0,j}$ the bare Rabi frequencies, which might not be the same on all ions.
The higher the motional energy is, the slower the Rabi oscillations will be and the faster they will dephase. This effect is clearly visible in the Rabi oscillations shown  in Fig.~\ref{fig:Rabiflops1d} after Doppler cooling (a) and polarization gradient cooling (b). While a single Rabi flop is barely visible for Doppler-cooled ions, the oscillations persist for many periods after polarization gradient cooling. As $\eta\sim \omega^{-1/2}$, carrier Rabi oscillations mostly probe the occupation of low-frequency modes. The bigger Lamb-Dicke parameter of low-frequency modes also explains that ions at the ends of the string experience a stronger Rabi frequency reduction because the motion of these ions has a stronger overlap with low-frequency modes (see also Fig.~\ref{fig:22ionspectrum1d}). It should be noted that the observed variation of the average Rabi frequency from ion to ion cannot be explained by effects such as the spatial profile of the laser beam or coupling strength variations by axial micromotion.  

A quantitative characterization of polarization gradient cooling is achieved by measuring the mean phonon numbers of individual collective modes by sideband spectroscopy. With the exception of the center-of-mass mode, it is very hard to extract mean phonon numbers from sideband measurements of collectively excited ions, such as in Fig.~\ref{fig:22ionspectrum1d}. We therefore use our single-ion addressing capability to transfer all ions except one to a different Zeeman state so that the axial laser beam at 729~nm can be used to drive sideband transitions coupling to a single ion only. For a given mode $k$ of interest, the ion with the largest Lamb-Dicke parameter $\eta_{jk}$ is chosen in order to minimize the probe pulse length.

\begin{figure}[ht]
    \centering
\includegraphics[scale=1.2]{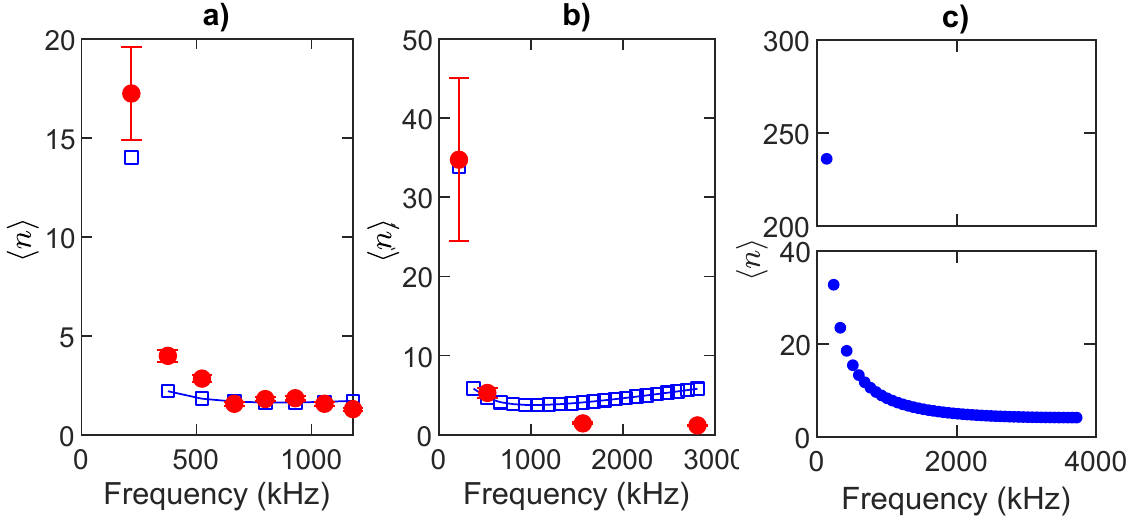}
    \caption{Cooling of linear ion strings. Mean phonon numbers for various motional modes for (a) an 8-ion, (b) a 22-ion, and (c) a 51-ion string. The red circles are the experimentally measured $\braket{n}$, obtained by single-ion sideband spectroscopy. Blue squares are experimentally estimated values of $\braket{n}$ obtained by fitting carrier Rabi oscillations on all ions with the model discussed in the main text. For the 51-ion case, individual ion interrogation was not possible due to technical limitations. Calibration of the intensity of the polarization gradient indicates that $\xi=\sqrt{5/6}$ is reached at oscillation frequencies of about 475~kHz (8 ions), 1225~kHz (22 ions) and 1220~kHz (51 ions).}
    \label{fig:8-22ionnbarvsfreq}
    

\end{figure}

Figure~\ref{fig:8-22ionnbarvsfreq} (a) shows measurements of the mean phonon numbers of all axial motional modes of an eight-ion string after polarization gradient cooling. Panel (b) shows similar measurements for four motional modes in a 22-ion string. With the exception of the center-of-mass mode, all motional modes are cooled down to a few motional quanta. We attribute the rather high number of quanta in the center-of-mass mode to a competition of laser cooling with motional heating by electric field noise. Given the measured heating rate of 1350(85)~quanta/s for a single ion, we expect the heating rate by electric field noise to increase by a factor of $N$ (as the electric field noise seen by different ions is strongly correlated), leading to about 11000~quanta/s for an 8-ion and 30000~quanta/s for a 22-ion string. These heating rates are comparable to the rate of heating that is intrinsic to the laser cooling scheme; in contrast, for a single ion, heating by electric field noise is substantially reduced and should therefore not affect the polarization-gradient cooling limit. Moreover, the measurement might slightly overestimate the cooling limit due to motional heating occurring within the time period (typically 250 $\mu$s) between the end of the laser cooling pulse and the start of the sideband probe pulse. While polarization gradient cooling cannot prepare the center-of-mass mode in very low phonon number states, it nevertheless substantially reduces the mode occupation after Doppler cooling that is estimated to be 158(18), 184(17) for ion strings with 8 and 22 ions, respectively.

Single-ion addressing is a technique that might not be available in every experiment with laser-cooled ion strings. For this reason, a semi-quantitative analysis of the cooling performance, that can be carried out with a laser exciting all ions equally, would be a useful tool. Towards this aim, we analyzed the excited state probabilities $p_j(t)$ of the $j^{th}$ ion when driving carrier Rabi oscillations for a thermal state  $\overline{\mathbf{n}}=(\braket{\mathbf{n}_1},...,\braket{\mathbf{n}_N})$,
\begin{equation}
    p_j(t) = \sum_\mathbf{n} p_{\overline{\mathbf{n}}}(\mathbf{n}) \sin^2(\Omega_j^{\mathbf{n}}t/2),
    \label{eq:RabiOscillations}
\end{equation}
which, through the dependence of $\Omega_j^{\mathbf{n}}$ on the motional state, provide information about the phonon distribution $p_{\overline{\mathbf{n}}}$ of all motional modes involved. As the damping and frequency shift of Rabi oscillations is predominantly caused by thermally populated low-frequency modes, one cannot hope to extract the mean phonon numbers of all motional modes from such a data set. Therefore, for the analysis of a polarization-gradient cooled ion string,  the mean phonon number $\braket{n_i}$, ($1\le i\le N$) of the individual modes oscillating at frequency $\omega_i$ are parametrized by a model with only three parameters,
\begin{equation}
  \braket{n_i}=\begin{cases}
    n_c, & \text{if $i$ = 1},\\
    \frac{n_0}{2}\left(\frac{ \omega_i}{\omega_0}+\frac{\omega_0}{ \omega_i}\right), & \text{if $i$ = 2 to $N$}, 
    \end{cases}
    \label{eq:myltiIonnbar}
\end{equation}
where $n_c$ is the mean phonon number of the center-of-mass mode, $n_0$ the lowest mean phonon number, and $\omega_0$ the frequency at which $n_0$ is reached. The choice of this model is motivated by the fact that for a single ion the observed mean phonon numbers conform sufficiently well to the functional dependence on the trap frequency predicted by eq.~(\ref{eq:nbarVsXi}). For an ion crystal, we expect the same dependence of the mean phonon number on the mode frequency, except for the center-of-mass mode which is affected by motional heating.

Fitting to the experimental data is computationally expensive as
expression (\ref{eq:RabiOscillations}) 
can be neither analytically calculated nor exactly evaluated numerically due to the enormous state space of the $N$ harmonic oscillators involved. For a given thermal state $\braket{\mathbf{n}}$, we therefore have to sample from the thermal distributions many times in order to arrive at a numerical estimate of $p_j(t)$. This presents an obstacle to using standard minimization algorithms in the data fitting routine as the fit function now has become a stochastical variable: every time we evaluate $p_j(t)$ for a given set of parameters, we have to sample from the thermal distributions and will obtain a slightly different function value. Therefore, we cannot use any gradient-based minimization routine; also, gradient-free routines such as the simplex algorithm can fail \cite{Vugrin:2005}. To overcome this problem, we used a dividing rectangles algorithm, developed in the context of variational quantum simulation (see \cite{Kokail2019} and references therein), for finding the parameter set ($n_c$,$n_0$,$\omega_0$) that optimally fits a given data set.  

Figure \ref{fig:8-22ionnbarvsfreq} shows mean phonon numbers obtained by fitting Rabi oscillations of polarization-gradient cooled ion strings of 8, 22, and 51 ions. When fitting the data, we take into account that axial micromotion caused by an axial q-parameter of $q_z\approx 0.0013$ reduces the bare Rabi frequencies $\Omega_{0,j}$ (c.f.~eq.(\ref{eq:RabiFrequency})) of the outermost ions with respect to the ones of the innermost ions by about 2\% (11\%) for an ion string with 22 (51) ions.
For the 8-ion Rabi oscillations, the extracted mean phonon numbers are in reasonably good agreement with the single-ion sideband spectroscopy measurements. For the 22-ion Rabi oscillations, the extracted mean phonon number of the center-of-mass mode and another low-frequency mode also agrees with the sideband measurements, whereas carrier fitting seems to overestimate the populations in the high-frequency modes, whose thermal occupation only weakly affects the shape of the carrier oscillations. At the time the measurement was carried out, we did not have the ability to carry out sideband spectroscopy with 51 ions. The 51-ion carrier oscillations indicate a strongly occupied center-of-mass mode, which is probably caused by a very high heating rate, as creating a 51-ion linear string necessitated dropping the axial trap frequency to 127~kHz. For a Doppler-cooled string, we estimated a mean phonon number of 760(100) for the axial center-of-mass mode.

\subsection{Cooling of planar ion crystals}\label{sub:2D}
\begin{figure}[ht]
    \centering
 \includegraphics[scale=0.8]{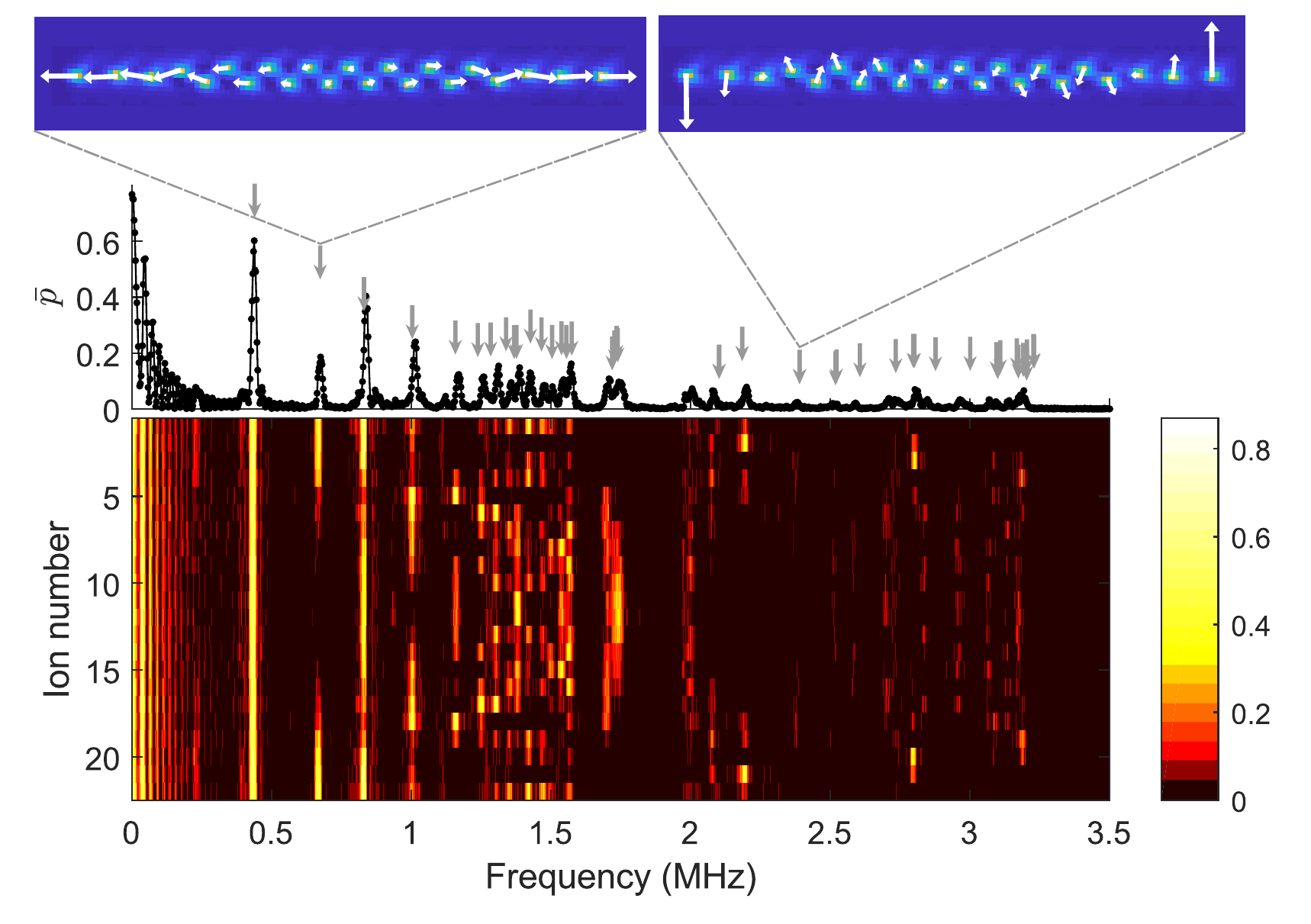}
    \caption{Single-ion-resolved spectrum and mean excitation of a 22-ion planar crystal after 1 ms polarization gradient cooling for $\omega_z = 2 \pi \times 438$~kHz, $\omega_x = 2 \pi \times 2.76$ MHz and $\omega_y = 2 \pi \times 2.518$ MHz. The crystals' mode frequencies calculated using a pseudopotential approximation approach agree well with the experimentally measured sideband resonances (indicated by grey arrows). 
    The mode vectors of two motional modes are shown on top of the spectrum (white arrows overlaid on CCD pictures of the ion crystal). The left mode is the equivalent of the stretch mode of a linear crystal. The strong axial amplitude of the outer ions results in a strong excitation of these ions by a laser probing the crystal. These two examples show that the calculated crystal's mode structure can explain the observed spatially resolved excitation pattern on the ions.}
    \label{fig:spectrum22ionsplanar}
    
\end{figure}

Operating our linear trap with a higher axial confinement and an increased frequency splitting of the transverse normal modes enables the creation of planar crystals in a plane that is normal to the direction having the highest frequency. For center-of-mass mode frequencies of $\omega_z = 2\pi \times 438$~kHz, $\omega_x = 2 \pi \times 2.76$ MHz and $\omega_y = 2 \pi \times 2.51$ MHz, we were able to create the 22-ion zig-zag crystals displayed in Fig.~\ref{fig:spectrum22ionsplanar}. This trap configuration opens up the possibility of cooling the in-plane modes of motion in an axial polarization gradient without suffering from micromotion-induced line broadening that those ions experience which are not located directly on axis.

\begin{figure}
\centering
\includegraphics[scale=.75]{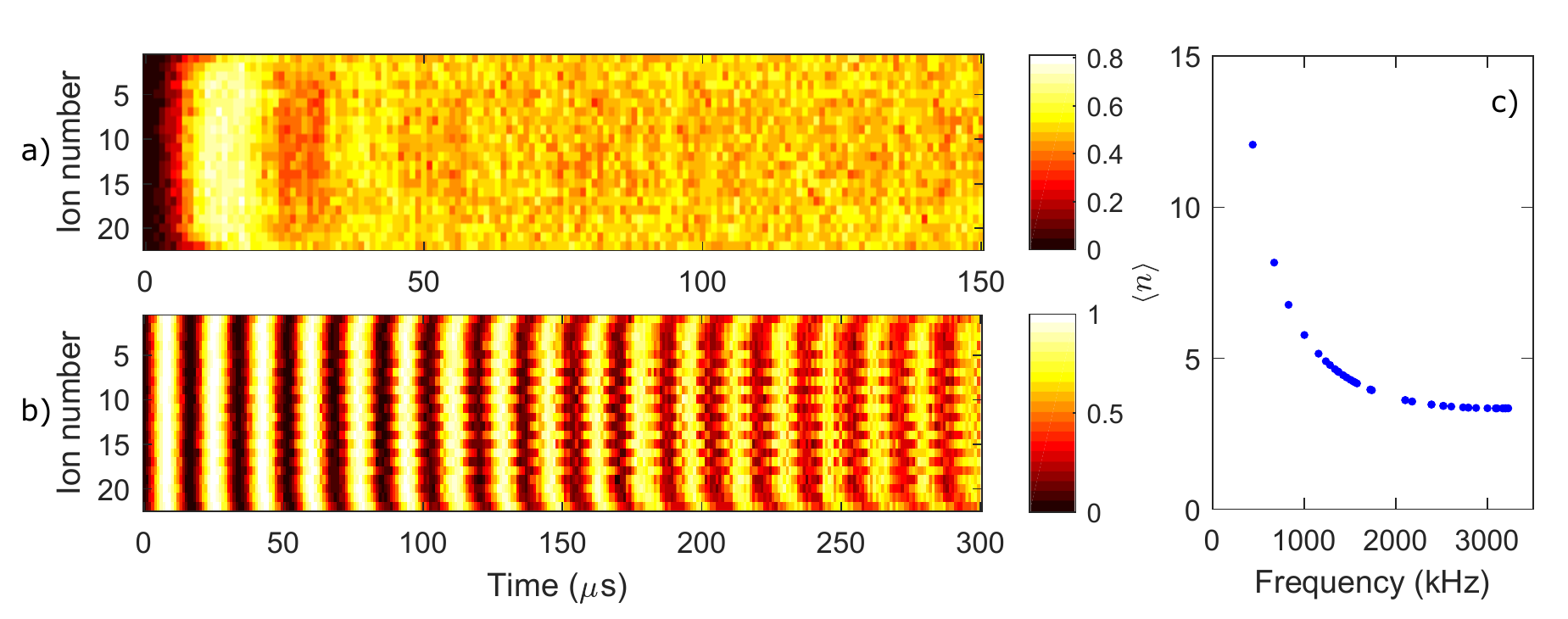}
\caption{Carrier Rabi oscillations of a 22-ion planar crystal driven by a 729~nm laser beam propagating along the symmetry axis of the linear trap after Doppler cooling (a) and polarization-gradient cooling (b), which dramatically improves the quality of the oscillations. (c) Estimated mean phonon numbers for all motional modes obtained by fitting these carrier oscillations with the model described by eqs.~(\ref{eq:RabiOscillations}) and (\ref{eq:myltiIonnbar})}. 
\label{fig:Planar22RabiNbars}
\end{figure}

Fig.~\ref{fig:spectrum22ionsplanar} shows a spectrum of the blue sideband transitions of the planar modes, obtained by polarization-gradient cooling the crystal for 1~ms and mapping motional state excitation to the $D_{5/2}$ state by a sideband pulse with a 729~nm laser beam propagating along the axial direction. Here, the 2D-plot shows the spatially resolved excitations of all the ions, which are numbered according to their horizontal position in the crystal. The average excitation $\overline{p}$ is displayed on top of the plot as a black line. We were able to identify sideband resonances (grey arrows) of 40 out of the 44 in-plane collective modes of motion. The measured sideband frequencies are mostly in good agreement with mode frequency calculations based on both pseudopotential theory and Floquet-Lyapunov theory \cite{Landa:2012} (for the frequencies of the in-plane modes, which we observe, the predicted values by the two approaches are pretty close). Moreover, we also see a good match between ions with the highest $D_{5/2}$ state excitation and the projection of their normal mode vector onto the axial direction. Two example sets of normal mode vectors (white arrows) are  graphically shown on top of images of the ion crystal.

A comparison of Rabi oscillations for a Doppler-cooled and a polarization gradient-cooled 2D-crystal is shown in Fig.~\ref{fig:Planar22RabiNbars}. Due to poor Doppler cooling results because of the micromotion-broadened cooling transition, no Rabi oscillations are visible in panel (a). In contrast, persistent Rabi oscillations are observed for polarization-gradient cooled ions. Fitting these oscillations with the model of eqs.~(\ref{eq:RabiOscillations}) and (\ref{eq:myltiIonnbar}), we estimate the center-of-mass mode's mean phonon number to be about 15 and all the other modes to be populated by eight or fewer phonons on average.

\section{Discussion and conclusion}\label{sub:con}
We have investigated polarization gradient cooling of long ion strings and planar crystals in a linear Paul trap. Tests with a single ion show that a simple cooling model for a moving polarization gradient predicts the cooling limit and cooling rate for experiments carried out deep in the LDR with reasonable accuracy. Moreover, we find an excellent match between experimental results and numerical simulations (see Fig.~\ref{fig:nbarvisXimovgrad}). At low frequencies, when the Lamb-Dicke parameter becomes big, we observe increased cooling limits and a shift of the optimum cooling conditions towards higher intensities of the lasers realizing the polarization gradient. 

With long ion strings, we observe multi-mode cooling over a wide range of frequencies, with mean phonon numbers significantly below the ones that can be achieved by Doppler cooling on the same transition. Towards the low-frequency end, and in particular for the center-of-mass frequency, the cooling limit predicted by the simple model is not reached; in addition, it appears that these modes are optimally cooled for saturation intensities that are higher than predicted. The reason for this behaviour is currently not well understood. As $\eta\sim N^{-1/2}$, the lowest-frequency mode has a much lower Lamb-Dicke parameter than a single ion would have. The presence of spectator modes might lead to an increased cooling limit and possibly to a reduced cooling rate; as a consequence, electric field noise might impede cooling of the center-of-mass mode. For the planar crystals that we investigated, nearly all of the $2N$ in-plane modes of the crystal had ions with sufficiently big Lamb-Dicke parameters along the trap axis to enable polarization-gradient cooling of these modes. This makes polarization gradient cooling a valuable tool for cooling many modes of the crystal below the Doppler limit.

Achieving sub-Doppler cooling at high rates is another asset of  polarization gradient cooling. For Doppler cooling, the rate at which phonons can be extracted scales as $W_{Doppler}\sim\eta^2\omega$ and therefore does not depend on the trap frequency $\omega$ (because $\eta\sim\omega^{-1/2}$). For polarization gradient cooling, we have $W_{PGC}\sim\eta^2\omega(\Gamma/\Delta)\xi^2$. When the depth of the optical potential is chosen to yield the lowest mean phonon number, i.e. at $\xi=\sqrt{5/6}$, the cooling rate achieved by PGC is expected to be smaller than the Doppler cooling rate if $\Gamma\ll \Delta$. On the other hand, one can achieve higher cooling rates for PGC than with Doppler cooling when $\xi>\sqrt{5/6}$ but at the expense of somewhat higher mean phonon numbers \cite{Wineland1992}. 

Another interesting application of polarization gradient cooling is sub-Doppler multi-mode cooling over a large frequency range. Polarization gradient cooling cannot compete with EIT cooling when it comes to cooling multiple modes close to the ground state \cite{Lechner2016,Lin:2013, Scharnhorst:2018}. EIT cooling, however, only achieves high cooling rates over a comparatively small frequency range due to the narrow dressed state facilitating the cooling. Moreover, since the highest EIT cooling rate scales as $W_{EIT}\sim\omega$ \cite{Morigi:2003}, EIT cooling becomes slow for low-frequency modes in contrast to polarization gradient cooling, for which the cooling rate can stay high at low confinement, where the state-dependence of the potential seen by the ion becomes more pronounced.

The experimental setup used for the measurements described in this paper is dedicated to carrying out quantum simulations with long ion strings via engineered spin-spin couplings, which are mediated by transverse motional modes cooled close to the ground state. While the axial modes of motion do not need to be cooled to very low quantum numbers, hot axial modes can nevertheless give rise to spurious couplings inducing decoherence in the spin-spin couplings, as well as to coupling strengths variations in addressed single-ion operations with a tightly focused beam. Earlier attempts failed to cool these modes using an EIT cooling setup, which was used for ground-state cooling the transverse modes \cite{Lechner2016}. In contrast, polarization gradient cooling has been demonstrated to be able to prepare the axial modes at lower phonon numbers than the ones achievable by Doppler cooling. For long strings with a very large ratio between the lowest and the highest collective mode frequency $\omega$, cooling range limitations imposed by $\langle n\rangle\sim \omega_0/\omega + \omega/\omega_0$ might become a problem. Here, use of concatenated cooling pulses in a polarization gradient with an optical depth decreasing over time might extend the cooling range in a way similar to what has been demonstrated with sideband and EIT cooling \cite{Monroe1995,Scharnhorst:2018}. 

\section{Acknowledgements} 
We thank Haggai Landa for sharing with us his code for normal mode calculations using the Floquet-Lyapunov approach. CR acknowledges useful discussions with Alex Retzker. 
MKJ would like to acknowledge Pavel Hrmo for discussions and careful reading of the manuscript. The project leading to this application has received funding from the European Research Council (ERC) under the European Union’s Horizon 2020 research and innovation programme (grant agreement No 741541), and from the European Union’s Horizon 2020 research and innovation programme under grant agreement No 817482. Furthermore, we acknowledge support by the Austrian Science Fund through the SFB BeyondC (F71) and funding by the Institut f\"r Quanteninformation GmbH.

\appendix 
\section{Numerical simulations of polarization gradient cooling by the Lindblad master equation approach}\label{sub:ap:num} 
We model the ion as a four-level atom that is driven on the $|S_{1/2},m_{j}=\pm 1/2\rangle\leftrightarrow|P_{1/2},m_{j}=\mp1/2\rangle$ transitions by two classical light fields. Our numerical simulations are based on solving the Lindblad master equation,
\begin{equation}
\frac{d\hat{\rho}\bigl(\mathrm{t}\bigr)}{dt}=[H,\hat{\rho}]+\mathcal{L}_{diss}\bigl[\hat{\rho}(t)\bigr],\label{eq:Lindblad-equation}
\end{equation}
where $\rho$ is the density matrix of the ion trapped in a one-dimensional harmonic potential whose state space is truncated to the lowest $n_{max}$ Fock states. The Hamiltonian $H=H_{trap}+H_{al}$ consists of two parts, with $ H_{trap}=\hbar\omega_z(a^\dagger a+\frac{1}{2})$ describing the ion's motion in the trap, where $a^\dagger$ and $a$ denote the motional raising and lowering operators, and the atom-light interaction Hamiltonian $H_{al}$. In the frame rotating at the laser detuning $\Delta$ from the atomic transition, it reads 
\begin{align}
H_{al}  = & \frac{\Omega_{\text{Str}}}{2}\sqrt{\frac{1}{3}}\left((e^{-i\left(\phi-\frac{\pi}{4}\right)}e^{-ik\hat{z}}+e^{i\left(\phi-\frac{\pi}{4}\right)}e^{ik\hat{z}})\sigma^{-}_{\text{cir}+}+h.c.\right)\nonumber\\
+&\frac{\Omega_{\text{Str}}}{2}\sqrt{\frac{1}{3}}
\left((e^{-i\left(\phi+\frac{\pi}{4}\right)}e^{-ik\hat{z}}+e^{i\left(\phi+\frac{\pi}{4}\right)}e^{ik\hat{z}})\sigma^{-}_{\text{cir}-}+h.c.\right)\nonumber\\
-&\Delta\left(\sigma^{+}_{\text{cir}+} \sigma^{-}_{\text{cir}+}+\sigma^{+}_{\text{cir}-}\sigma^{-}_{\text{cir}-}\right).
\end{align}
where $\sigma^{+}$ and $\sigma^{-}$ are the raising and lowering operators on the transitions driven by the two circularly polarized standing waves, each of which is described by two counterpropagating beams with wave vectors $\pm k\mathbf{e}_z$. The symbol $z$ denotes the position operator, which we express as $k\hat{z}=\eta(a+a^\dagger)$. Furthermore, $\Omega_{\text{str}}$ is the Rabi frequency of the stretched transition $|S_{1/2},m_{j}=+1/2\rangle\leftrightarrow|P_{3/2},m_{j}=+3/2\rangle$ and $\phi$ is the laser phase at the ion position. We do not model the frequency shifts of the atomic states in a non-zero magnetic field, which we assume to be small in comparison to $\Delta$. 

The Liouvillian operator $\mathcal{L}_{diss}$ accounts for spontaneous decay during the laser-ion interaction. It is constructed using the jump operator describing the decay process from the required transitions, which reads as
\begin{equation}
L=\mathop{\mathop{\sum_{m}\hat{J}_{m}\hat{\rho}J_{m}^{\dagger}-\frac{1}{2}\mathop{\sum_{m}\big(\hat{J}_{m}^{\dagger}\hat{J}_{m}\hat{\rho}+\hat{\rho}\hat{J}_{m}^{\dagger}\hat{J}_{m}}\big)}},
\end{equation}
where  $J_{m}=\sum_{q}p_{mq}C_{m}\Gamma(e^{-ik_{q}z}\sigma_{m}^{-})$. Here $m$ stands for an index indicating transitions associated with 4 decay channels that exist between the Zeeman levels of the excited state and the ground state. We take into account the spatial dipole radiation pattern by subdividing each decay channel into three terms which transfer a momentum of $\hbar k_q$ with $k_{q}\in\{-k,0,+k\}$ to the ion along its direction of oscillation \cite{Molmer:1993}. The $C_{m}$ are the Clebsch-Gordan coefficients, which
are $\sqrt{1/3}$ and $\sqrt{\ensuremath{2/3}}$ for $\Delta m_{j}=0$ and $\pm1$, respectively. The values of the pre-factors $p_{mq}$ are given in table \ref{tab:tablepmq}.
\begin{table}[h!]
\caption{Table showing $p_{mq}$ for jump operators defined earlier.}
\label{tab:tablepmq}
\centering
\begin{tabular}{|l|l|l|l|}
\hline
 $p_{mq}$& $m=+1$  & $m=0$  & $m=-1$  \\ \hline
 $q=+1$ & $\sqrt{1/5}$ & $\sqrt{1/10}$ &  $\sqrt{1/5}$  \\ \hline
 $q=0$ &  $\sqrt{3/5}$ & $\sqrt{8/10}$ &  $\sqrt{3/5}$ \\ \hline
 $q=-1$ & $ \sqrt{1/5}$ &  $\sqrt{1/10}$ &  $\sqrt{1/5}$ \\ \hline
\end{tabular}
\end{table}

\section{Master equation simulations of cooling rate and cooling limit}
\label{sub:ap:MasterEquSim} 
\begin{figure}
\centering
\includegraphics[scale=0.7]{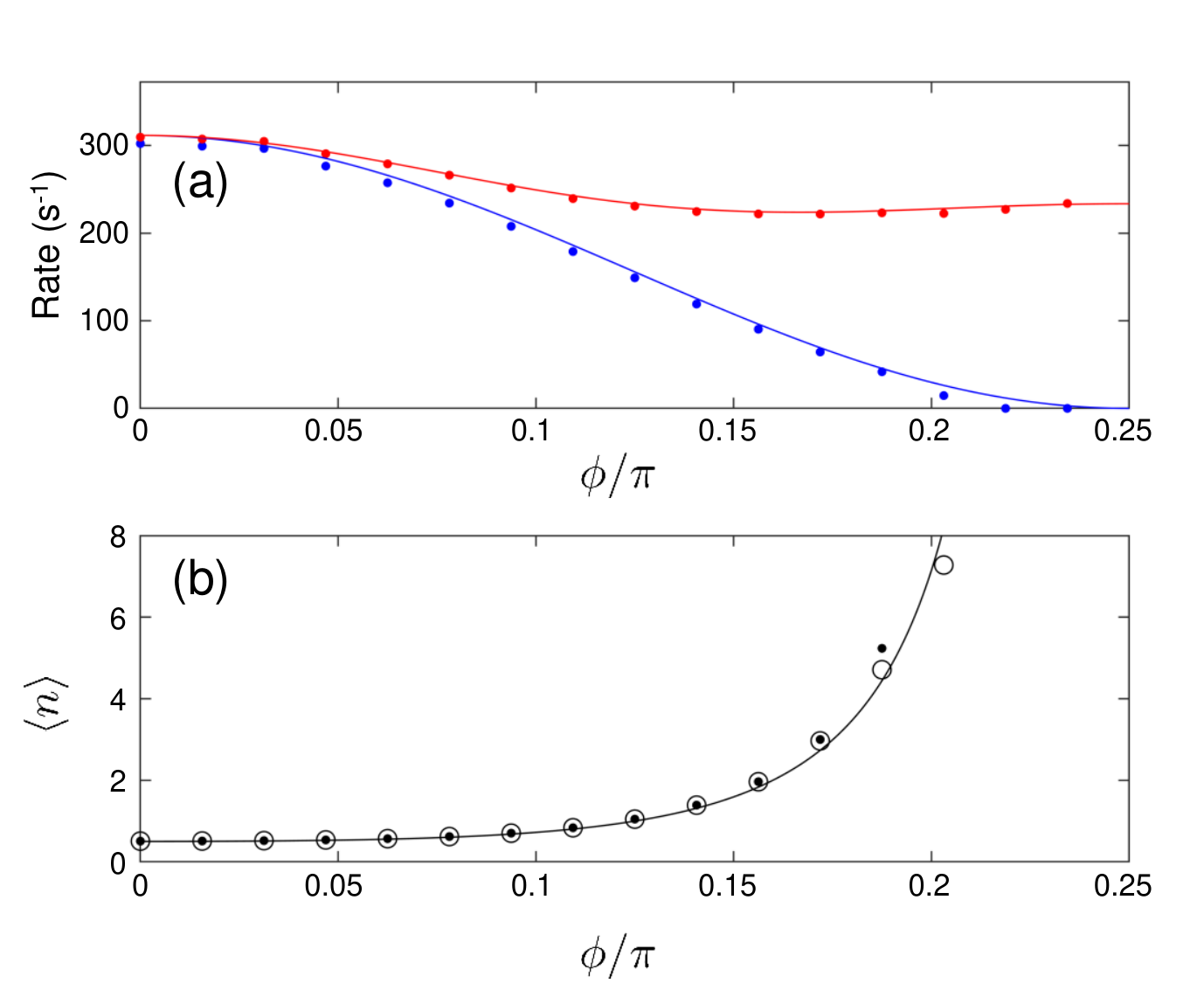}
\caption{Comparison of the predictions of the simple cooling model (solid lines) and master equation simulations (filled symbols) for different positions within the polarization gradient. (a) Cooling rate (blue) and heating rate (red). (b) Average phonon number in steady state. Note that the rates are reduced by a factor of 100, compared to the experiment, because of the increase in transition wavelength ($\lambda\rightarrow 10\times\lambda$).}
\label{fig:HeatingCoolingMasterSim}
\end{figure}

We validate the predictions of the simple cooling model presented in section~\ref{sub:theory} by comparison with master equation simulations. For a numerical determination of the cooling rate and cooling limit, we time-evolve an ion initially prepared in the motional ground state under eq.~(\ref{eq:Lindblad-equation}) until the motional state reaches its steady state. Figure~\ref{fig:HeatingCoolingMasterSim} shows the results for a Ca$^+$ ion oscillating at $\omega_z=2 \pi \times 1088$~kHz for different phases $\phi$ of the polarization gradient. In order to achieve a comparison deep inside the LDR, we artificially increase the ion's transition wavelength by a factor of 10 (so that $\eta=0.017$). As a consequence, the dynamics is slowed down a by a factor of 100. For determining the heating rate $H(\phi)$, we fit a tangent at $t=0$ to the mean phonon number $\braket{n(t)}$ using a motional state space comprising the lowest $m=24$ Fock states. For determining the steady-state phonon number and the cooling rate, we fit the data by a function $f(t)=n_m(1-\exp(-W_m t))$. For phases $\phi$ approaching $\pi/4$, the extracted steady-state phonon number and also the cooling rate depend on the state space cut-off $m$. Therefore, we carried out these simulations for $m\in\{4,5\dots 24\}$ and used a finite-size scaling to extrapolate $n_m$ to infinity to obtain the true steady-state mean phonon number $n_\infty$. There are indications that this procedure slightly overestimates $n_\infty$. In Fig.~\ref{fig:HeatingCoolingMasterSim}(b), $n_\infty(\phi)$ is indicated by filled symbols whereas the open symbols represent $n_{m=24}(\phi)$. The cooling rate $W(\phi)$ is extracted by combining the results for $H(\phi)$ with $n_\infty$. 

\begin{figure}
\centering
\includegraphics[scale=0.7]{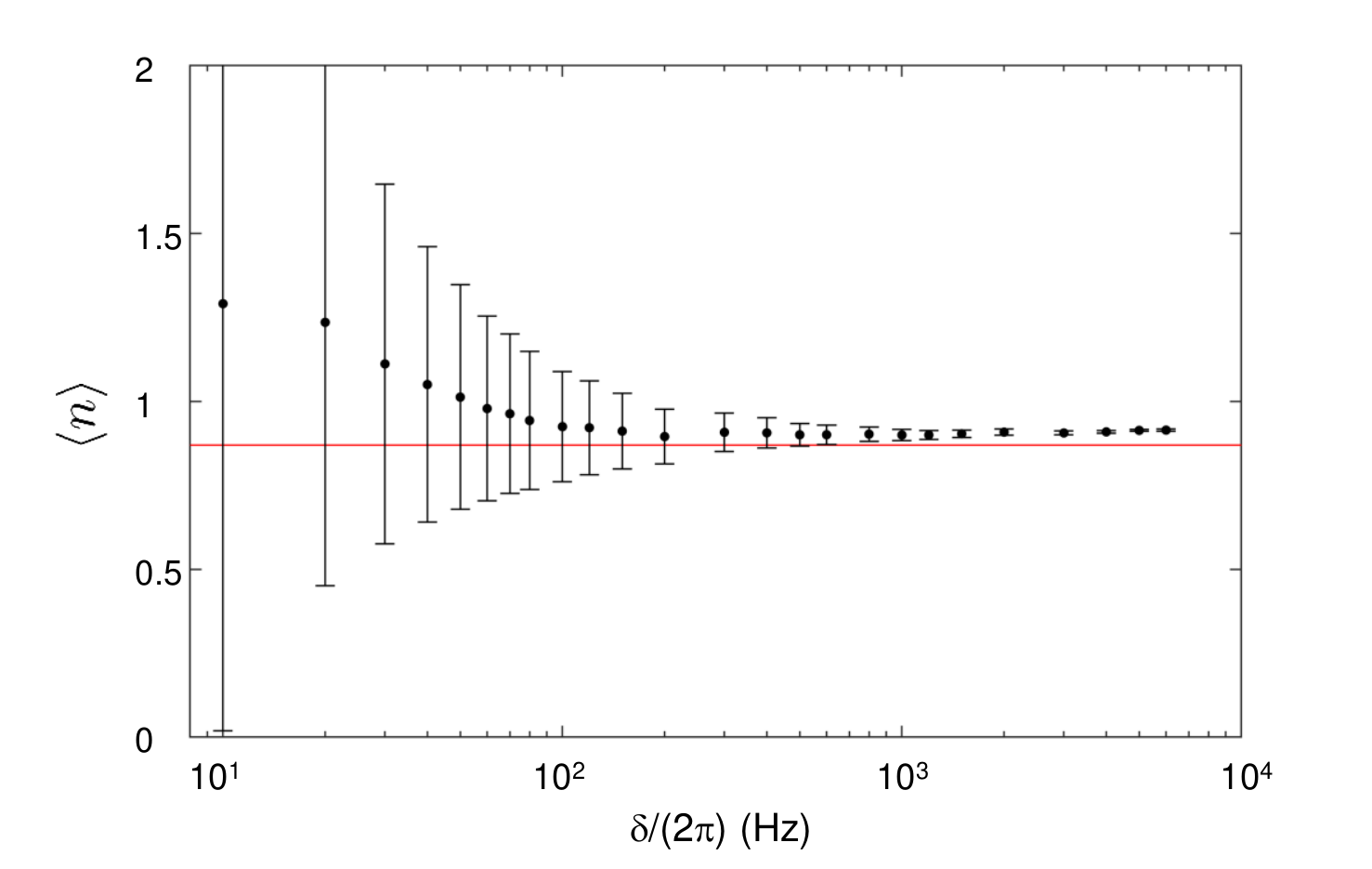}
\caption{Simulated mean phonon numbers (black dots) of a single ion as a function of frequency detuning $\delta$ between the two counter-propagating laser beams creating a moving polarisation gradient at the ion position. The error bar indicate the width of the distribution of mean phonon numbers that result from stopping the cooling pulse at a random phase of the moving gradient. The prediction of the simple cooling model is indicated by the red line. Again, for a comparison with the experiment, the horizontal axis should be multiplied by a factor of 100.}
\label{fig:GradientSpeedMasterSim}
\end{figure}

In a second investigation shown in Fig.~\ref{fig:GradientSpeedMasterSim}, we looked into the cooling limit in a moving polarisation gradient and its dependence on the relative frequency detuning $\delta$ between the two counter-propagating beams. The simple model predicts a constant cooling limit provided that $\delta$ is considerably bigger than the cooling rate. Again, we set $\omega_z=2\pi\times 1088$~kHz, but replace $\eta\rightarrow \eta/10$ in order to be deep in the LDR. The data shown in Fig. ~\ref{fig:GradientSpeedMasterSim} were obtained for optimum cooling conditions, $\xi=\sqrt{5/6}$ using a state space comprising the twenty lowest Fock states. After temporally averaging the steady-state $n(t)$ over a period $\tau=2\pi/\delta$, we find a good agreement between the master equation simulation and the simple model for sufficiently fast values of $\delta$. If $\delta$ becomes smaller than the cooling rate, $\braket{n}$ goes up and the variance of $n(t)$ (indicated by the error bars in the figure) strongly increases.

\section{Comparison of master equation simulations with experimental data} \label{sub:ap:comparison} 
\begin{figure}
\centering
\includegraphics[scale=0.7]{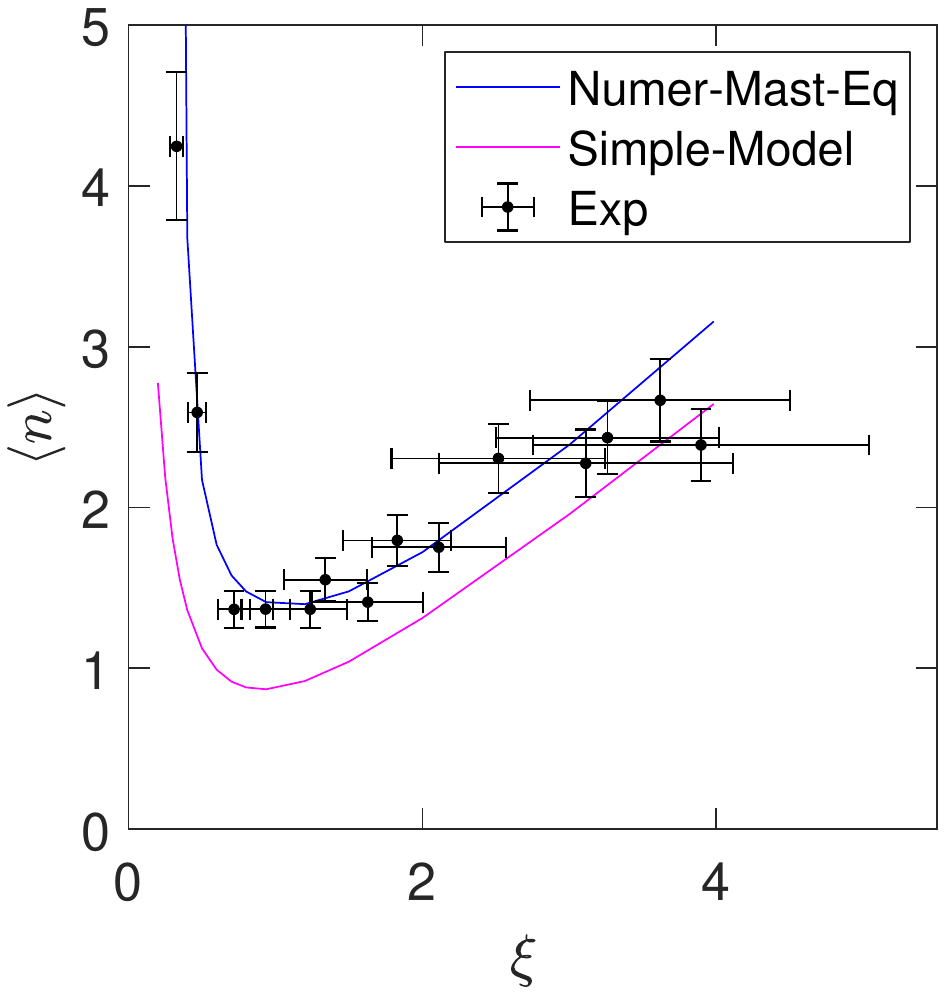}
\caption{Cooling limit as a function of $\xi =(\Delta s/(3 \omega_z))$. The blue curve represents the simulation results with the master equation and the magenta one represents the results from the semi-classical treatment. }
\label{fig:nbarvisXimovgrad}
\end{figure}

The motional energy of a single ion shown as red data points in Fig.~\ref{fig:SingleIonMeasurement} of the main text are slightly higher than the prediction of the simple cooling model, which is shown in Fig.~\ref{fig:nbarvisXimovgrad} as a pink line. Numerical simulations of the cooling limit (blue curve) agree well with the experimental data and show that this discrepancy can be explained by the fact that the ions are not deep in the LDR ($\eta=0.17$). For small values of $\xi$, the simulation shows non-thermal phonon number distributions with tails that are more heavily occupied than in a thermal state. For this reason, we carried out the simulation once more for motional state spaces of different size including up to 24 Fock states and used a finite size scaling for extracting the mean phonon number.

\section*{References}

\providecommand{\newblock}{}

\end{document}